\documentclass{aip-cp}

\usepackage[numbers]{natbib}
\usepackage{rotating}
\usepackage{graphicx}
\usepackage{siunitx}
\DeclareSIUnit\electron{e^{-}}
\usepackage{textgreek}

\usepackage{tikz}
\newcommand\copyrighttext{%
  \footnotesize This article may be downloaded for personal use only. Any other use requires prior permission of the author and AIP Publishing. The following article appeared in AIP Conf. Proc. \textbf{2054}, 060077 (2019) and may be found at https://doi.org/10.1063/1.5084708 .}
\newcommand\copyrightnotice{%
\begin{tikzpicture}[remember picture,overlay]
\node[anchor=south] at (current page.south) {\fbox{\parbox{\dimexpr\textwidth-\fboxsep-\fboxrule\relax}{\copyrighttext}}};
\end{tikzpicture}%
}

\usepackage{tikz}
\usetikzlibrary{shapes,arrows}
\tikzstyle{every node}=[font=\footnotesize]
\definecolor{slacred}{RGB}{140, 21, 21}
\tikzstyle{param} = [rectangle, text centered, node distance=2.5em]
\tikzstyle{io} = [trapezium, trapezium left angle=70, trapezium right angle=110,  line width=0.5pt, minimum height=2em, text width=3.2em, text centered, draw=black, node distance=4.5em]
\tikzstyle{block} = [rectangle, line width=0.5pt, draw, text width=3.5em, text centered, minimum height=2.3em, node distance=3.8em]
\tikzstyle{pblock} = [rectangle, line width=0.5pt, draw, text width=4em, text centered, minimum height=2.3em, node distance=6em]
\tikzstyle{weight} = [rectangle, line width=0.5pt, draw=slacred, text width=3.0em, text centered, rounded corners, minimum height=2em, node distance=3.8em]
\tikzstyle{line} = [draw, line width=1pt, -latex']
\tikzstyle{op} = [draw, circle, line width=0.5pt, node distance=3.8em]%, minimum height=1em]

\makeatletter 
\renewcommand\@biblabel[1]{#1.} 
\makeatother

\begin{document}

% Title portion
\title{Ultrafast Processing of Pixel Detector Data\\with Machine Learning Frameworks}

\author[aff1]{G.~Blaj\corref{cor1}}%\noteref{note1}}%\noteref{note1,note2}}
\author[aff1]{C.-E.~Chang}
\author[aff1]{C.~J.~Kenney}

%\eaddress[url]{http://www.aip.org}
%\author[aff2,aff3]{Author's Name\noteref{note2}}
%\eaddress{anotherauthor@thisaddress.yyy}

\affil[aff1]{SLAC National Accelerator Laboratory, 2575 Sand Hill Road, Menlo Park, CA 94025, U.S.A.}
%\affil[aff2]{Additional affiliations should be indicated by superscript numbers 2, 3, etc. as shown above.}
%\affil[aff3]{You would list an author's second affiliation here.}
\corresp[cor1]{Corresponding author: blaj@slac.stanford.edu}
%\authornote[note1]{SLAC-PUB-17277}
%\authornote[note2]{This is an example of second authornote.}

\maketitle
\copyrightnotice

\begin{abstract}
Modern photon science performed at high repetition rate free-electron laser (FEL) facilities and beyond relies on 2D pixel detectors operating at increasing frequencies (towards \SI{100}{\kilo\hertz} at LCLS-II) and producing rapidly increasing amounts of data (towards TB/s). This data must be rapidly stored for offline analysis and summarized in real time for online feedback to the scientists. While at LCLS all raw data has been stored, at LCLS-II this would lead to a prohibitive cost; instead, enabling real time processing of pixel detector data (dark, gain, common mode, background, charge summing, subpixel position, photon counting, data summarization) allows reducing the size and cost of online processing, offline processing and storage by orders of magnitude while preserving full photon information. This could be achieved by taking advantage of the compressibility of sparse data typical for LCLS-II applications. Faced with a similar big data challenge a decade ago, computer vision stimulated revolutionary advances in machine learning hardware and software. We investigated if these developments are useful in the field of data processing for high speed pixel detectors and found that typical deep learning models and autoencoder architectures failed to yield useful noise reduction while preserving full photon information, presumably because of the very different statistics and feature sets in computer vision and radiation imaging. However, the raw performance of modern frameworks like Tensorflow inspired us to redesign in Tensorflow mathematically equivalent versions of the state-of-the-art, ``classical'' algorithms used at LCLS. The novel Tensorflow models resulted in elegant, compact and hardware agnostic code, gaining 1 to 2 orders of magnitude faster processing on an inexpensive consumer GPU, reducing by 3 orders of magnitude the projected cost of online analysis and compression without photon loss at LCLS-II. The novel Tensorflow models also enabled ongoing development of a pipelined hardware system expected to yield an additional 3 to 4 orders of magnitude speedup, necessary for meeting the data acquisition and storage requirements at LCLS-II, potentially enabling acquiring every single FEL pulse at full speed. Computer vision a decade ago was dominated by hand-crafted filters; their structure inspired the deep learning revolution resulting in modern deep convolutional networks; similarly, our novel Tensorflow filters provide inspiration for designing future deep learning models for ultrafast and efficient processing and classification of pixel detector images at FEL facilities. 

\end{abstract}

% Head 1
\section{INTRODUCTION}

Modern photon science was enabled by advances of 2D pixel detector technology in the last decades \cite{delpierre2014history}. After the initial success of hard X-ray free-electron laser (FEL) facilities the last decade, two high repetition rate facilities are under development: EuXFEL (\SI{27}{\kilo\hertz}) \cite{koch2013detector} and LCLS-II (\SI{100}{\kilo\hertz}) \cite{schoenlein2015new}. Ultrafast data acquisition and processing will be critical in multiple applications: for online monitoring, data reduction (photon counting), sparsification, storage, and offline analysis.

A typical LCLS large area detector \cite{hart2012cspad} (2.2 Mpixel/frame, 2 bytes/pixel, \SI{120}{\hertz}) produces $\sim$\SI{0.5}{GB/s} and all raw data is saved, resulting in tens of petabytes of data requiring storage and analysis for the first 8 years of operation of LCLS \cite{blaj2014detector,blaj2015xray,thayer2016data}. Similar detectors operating at \SI{100}{\kilo\hertz} in LCLS-II will produce a massive \SI{430}{GB/s} \cite{blaj2015future}. The accelerator and detector \cite{blaj2015future} developments to meet this high rate are well underway. The data transfer and storage requirements can be met with multiple parallel lanes and larger versions of the existing data acquisition (DAQ) system \cite{herbst2014design}.

Raw images include noise, leading to inefficient lossless compression (typically a factor of $\approx 2$); typical LCLS-II images will be sparse, allowing a much higher compression rate after photon counting. If image processing (photon counting and sparsification) can take place in real time, sparse photon data is much easier to compress and use in online and offline analysis, resulting in dramatically lower requirements on real time storage and data analysis. For this reason, the prohibitive cost of storage and analysis of raw data at LCLS-II led to the development of a first version of the LCLS-II Data Reduction Pipeline \cite{thayer2016data}. However, an impractically large number of servers would still be required for real time reduction of 2D pixel detector data at \SI{100}{\kilo\hertz}.

Another area with rapid development in the last decade, driven by large increases of available data sets and stringent speed and efficiency requirements, is machine learning (especially computer vision applications). Several frameworks have been developed to allow scaling and optimization of system performance, leveraging advances in GPU technology while hiding system complexity, enabling simultaneously ultra rapid prototyping and efficient production systems; one such framework is Tensorflow, used for tensor (i.e., n-dimensional array) computation on CPUs and GPUs. Tensorflow allows a high level of abstraction, with compact, device-agnostic code, and orders of magnitude higher performance on GPUs compared to CPUs.

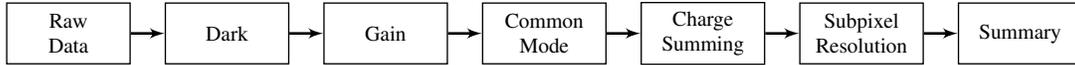
\begin{figure}[t]
  %\centerline {%
    \begin{tikzpicture}[node distance = 5.2cm, auto]
        % Place nodes
        \node [pblock] (raw)                {Raw\\Data};
        \node [pblock, right of=raw] (dg)   {Dark};
        \node [pblock, right of=dg] (gain) {Gain};
        \node [pblock, right of=gain] (cmc)   {Common Mode};
        \node [pblock, right of=cmc] (phot) {Charge Summing};
        \node [pblock, right of=phot] (sp)  {Subpixel Resolution};
        \node [pblock, right of=sp] (sum)   {Summary};
        % Draw edges
        \path [line] (raw) -- (dg);
        \path [line] (dg) -- (gain);
        \path [line] (gain) -- (cmc);
        \path [line] (cmc) -- (phot);
        \path [line] (phot) -- (sp);
        \path [line] (sp) -- (sum);
    \end{tikzpicture} 
  %}
  \caption{Typical steps in processing data from X-ray pixel detectors. Usually, Photon Finding and Subpixel Position are not extracted due to the high computational complexity, however, extracting them greatly increases data quality, enabling sparsification and efficient compression without photon loss. In this paper, we present complete Tensorflow models for all steps.}
  \label{fig:a}
\end{figure}

We present Tensorflow versions of usual, state-of-the-art algorithms used at LCLS for all steps of processing images from X-ray integrating pixel detectors and reducing it to photons. Due to the short pulse duration at FELs (in the order of femtoseconds), photon counting detectors can't be used; instead, integrating detectors must be used \cite{graafsma2015integrating}. Raw data from integrating pixel detectors requires (or at least benefits from) the following processing steps (Fig.~\ref{fig:a}): dark subtraction, common mode correction, gain (mode) calibration, single photon reconstruction (charge summing, subpixel resolution), and summarization (including histogramming, photon counting and sparsification, and data aggregation). We then compare their programming complexity and performance on both CPU and GPU with usual solutions using compiled Numba code. 

Using inexpensive consumer GPUs, we show 2 orders of magnitude improvement in data processing speed compared to ``classical'' CPU code (or 3 orders of magnitude improvement in cost for the same speed), already enabling real time processing and high rates of lossless compression for sparse data at LCLS with modest hardware costs. The approach presented here covers only 3 of the 6 orders of magnitude cost reduction required to match the LCLS-II speed requirements; however, it enables the development of an optimized, pipelined hardware approach which turns a prohibitively expensive task in a merely challenging one.

\section{METHODS}

Similar to the big data challenges of upcoming high repetition rate FELs were the challenges in computer vision a decade ago, which led to the current explosion of machine learning (ML) hardware and software and ongoing rapid development. One of the software frameworks for machine learning is Tensorflow, presenting a number of compelling advantages: Native Python support and trivial installation; compact, elegant code, enabling rapid development; hardware agnostic code enabling scalable deployment; native support for e.g., GPUs enabling high performance (requires the Nvidia CUDA Deep Neural Network library). Many other frameworks exist now.

Note that existing machine learning approaches to computer vision are not directly applicable to X-ray photon detection due to very different features, signal and noise statistics; while, e.g., autoencoders or deep learning models currently lose photon information, the models presented here preserve full photon information. For this reason, we developed novel low-level filters for all steps required to process data from 2D X-ray pixel detectors, which are mathematically equivalent to the current state-of-the-art algorithms for photon processing in 2D integrating pixel detectors at FELs.

To evaluate the performance of our Tensorflow models in challenging conditions, we intentionally maximized charge sharing \cite{blaj2017analytical} by (1)~using a Hammerhead ePix100 camera \cite{blaj2018hammerhead}, built with a novel sensor with a thickness of \SI{500}{\micro\metre} and high aspect ratio pixels of \SI{25x100}{\micro\metre}, (2)~operating the camera with the sensor underbiased (at \SI{100}{\volt}, just over the minimum bias for full depletion at \SI{90}{\volt}), and (3)~using relatively soft X-rays at \SI{5.9}{\kilo\electronvolt} generated with an $^{55}$Fe radioactive source. The Hammerhead ePix100 camera benefits from the high signal to noise ratio of ePix100 cameras \cite{blaj2015xray}. We acquired a large data set of \num{50000} frames (each containing \num{1}~MB of data from \num{176x3072}~pixels, equivalent to the typical ePix100 camera module of \num{704x768}~pixels).

We used two different computers: (1)~CPU system: Macbook Pro, 2.9 GHz Intel Core i7, 16 GB 2133 MHz LPDDR3, Numba 0.38, Tensorflow 1.1; and (2)~GPU system: Nvidia GTX 1060, Intel Xeon X3480, DDR3-1066 CAS latency 7, Tensorflow 1.4. We ran the Tensorflow models on both the CPU and GPU systems, comparing the performance with compiled Numba code (carefully optimized) running on the GPU system. The processing times we measured exclude the time to read data files, similar to real time data acquisition systems (receiving data from detectors over multiple \num{10}~Gbps links simultaneously in real time).
%
%i think sess.run routes data like this:
%
%feed: ssd -> dram -> gpu
%fetch: gpu -> dram
%
%
%btw i should've noticed this.. the following is not timed, but could've been part of tf model, and so timed
%
%hist0all += hist0
%histall += hist
%framesall += frames.sum(0)
%
%
%there are still two lines that i don't know if tf can do:
%
%allidx.append(idx)
%allphotons.append(photons)
%
%indeed, but I think it's a small overhead; also, timing points were excluding the buffering and the compiling of the results -

\section{NOVEL TENSORFLOW MODELS}
\subsection{Dark and Gain}
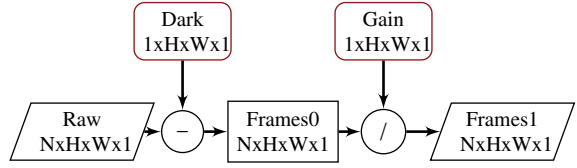
\begin{figure}[t]
  %\centerline {%
    \begin{tikzpicture}[node distance = 5.2cm, auto]
        \node [io] (raw)                {Raw\\NxHxWx1};
        \node [op, right of=raw] (op1)   {$-$};
        \node [weight, above of=op1] (dark)   {Dark\\1xHxWx1};
        \node [block, right of=op1] (dc)   {Frames0\\NxHxWx1};
        %\node [param, right of=dc](placeholder)      {};
        %\node [io, right of=placeholder] (dcm)   {Frames1\\NxHxWx1};
        \node [op, right of=dc] (op2)   {$/$};
        \node [weight, above of=op2] (gain)   {Gain\\1xHxWx1};
        \node [io, right of=op2] (out)   {Frames1\\NxHxWx1};
        % Draw edges
        \path [line] (raw) -- (op1);
        \path [line] (dark) -- (op1);
        \path [line] (op1) -- (dc);
        \path [line] (dc) -- (op2);
        \path [line] (gain) -- (op2);
        \path [line] (op2) -- (out);
    \end{tikzpicture} 
  %}
  \caption{Tensorflow models for dark and gain correction are trivial. We provide raw data in ``Raw'' input, formatted in a rank 4 tensor with N frames (minibatch size), image size HxW=\num{704x768}~pixels, and ``feature size'' \num{1} (correspoding to pixel intensity). The dimensions of each tensor are indicated in the diagrams. After dark subtraction and gain correction, we obtain the corrected minibatches Frames0 and Frames1, respectively. Accomodating multiple darks and gains in auto-ranging detectors is also straightforward (not shown here).}
  \label{fig:b}
\end{figure}

While dark and gain corrections are trivial \cite{blaj2016xray}, we present diagrams in Fig.~\ref{fig:b} to introduce the basics of tensor computation in Tensorflow. Stacks of frames called ``minibatches'' are submitted using rank 4 tensors: the first dimension, $N$, represents the number of frames in a minibatch; second and third dimensions represent the image size $H$~x~$W$, where $H$ and $W$ are the number of rows and columns, respectively. The last dimension represents a ``feature'' size (which is \num{1} for monochrome images, \num{3} for red-green-blue images); features can also represent other properties, e.g., matching a list of particular shapes). Subtraction and division are trivial and Tensorflow broadcasts the singleton dimensions automatically (similar to, e.g., Numpy).

\subsection{Common Mode}
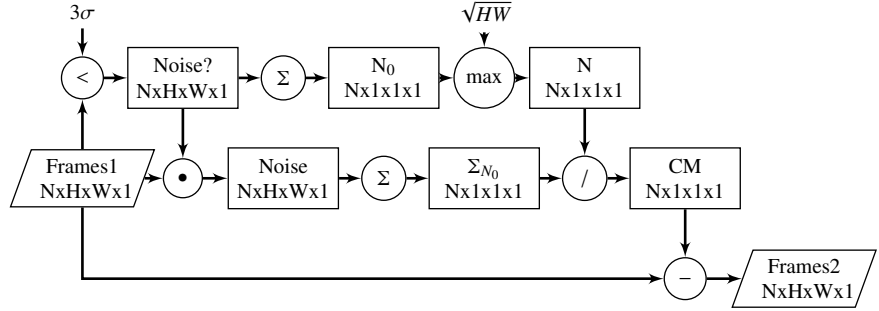
\begin{figure}[t]
  %\centerline {%
    \begin{tikzpicture}[node distance = 5.2cm, auto]
        % Place nodes
        \node [io] (in)                         {Frames1\\NxHxWx1};
        \node [op, above of=raw] (op1)          {$<$};
        \node [param, above of=op1](noise)      {$3\sigma$};
        \node [block, right of=op1] (test)      {Noise?\\NxHxWx1};
        \node [op, right of=raw] (op2)          {$\bullet$};
        \node [block, right of=op2] (innoise)   {Noise\\NxHxWx1};
        \node [op, right of=innoise] (op4)      {\textSigma};
        \node [op, right of=test] (op3)         {\textSigma};
        \node [block, right of=op4] (sumnoise)  {\textSigma$_{N_0}$\\Nx1x1x1};
        \node [block, right of=op3] (n0noise)   {N$_0$\\Nx1x1x1};
        \node [op, right of=n0noise] (op5)      {\footnotesize max};
        \node [param, above of=op5] (size)      {$\sqrt{HW}$};
        \node [block, right of=op5] (nnoise)    {N\\Nx1x1x1};
        \node [op, right of=sumnoise] (op6)     {$/$};
        \node [block, right of=op6] (cm)        {CM\\Nx1x1x1};
        \node [op, below of=cm] (op7)           {$-$};
        \node [io, right of=op7] (out)          {Frames2\\NxHxWx1};
        % Draw edges
        \path [line] (in) -- (op1);
        \path [line] (noise) -- (op1);
        \path [line] (op1) -- (test);
        \path [line] (test) -- (op2);
        \path [line] (in) -- (op2);
        \path [line] (op2) -- (innoise);
        \path [line] (test) -- (op3);
        \path [line] (innoise) -- (op4);
        \path [line] (op3) -- (n0noise);
        \path [line] (op4) -- (sumnoise);
        \path [line] (n0noise) -- (op5);
        \path [line] (size) -- (op5);
        \path [line] (op5) -- (nnoise);
        \path [line] (nnoise) -- (op6);
        \path [line] (sumnoise) -- (op6);
        \path [line] (op6) -- (cm);
        \path [line] (cm) -- (op7);
        \path [line] (in) |- (op7);
        \path [line] (op7) -- (out);
    \end{tikzpicture} 
  %}
  \caption{Tensorflow model for common mode correction appears complex, however, most of the diagram performs a more robust noise segmentation and average, which scales to high beam intensity without introducing spurious noise when only a small fraction of pixels sample the noise (within $3 \sigma$ of \num{0}); this design can be simplified at low photon occupancy (i.e., sparse photons).}
  \label{fig:c}
\end{figure}

The common mode model in Fig.~\ref{fig:c} appears complex, however, its operation is relatively simple: the pixels without signal (i.e., pixel signal smaller than $3 \sigma$) are segmented, the number of segmented pixels is calculated in the top branch, their total signal is calculated in the middle branch, and their ratio (i.e., the mean pixel background) is obtained and subtracted from the input signal \cite{blaj2016xray}. Note however the $\sqrt{H W}$ term which guarantees that the model works correctly also at high photon occupancy (when too few pixels measure the background and estimate a common mode correction based on only a few pixels, thus introducing spurious noise). If the camera is known to operate only at low photon occupancy, a much simpler and faster model can be used instead. While the model we present calculates the frame common mode, other types of common mode (row, column, block) can be easily extracted by reshaping tensors accordingly and summing only along the relevant dimensions.

\subsection{Photon Charge Summing}

\begin{figure}[!ht]
  \centerline {%
    \includegraphics[width =6.5in]{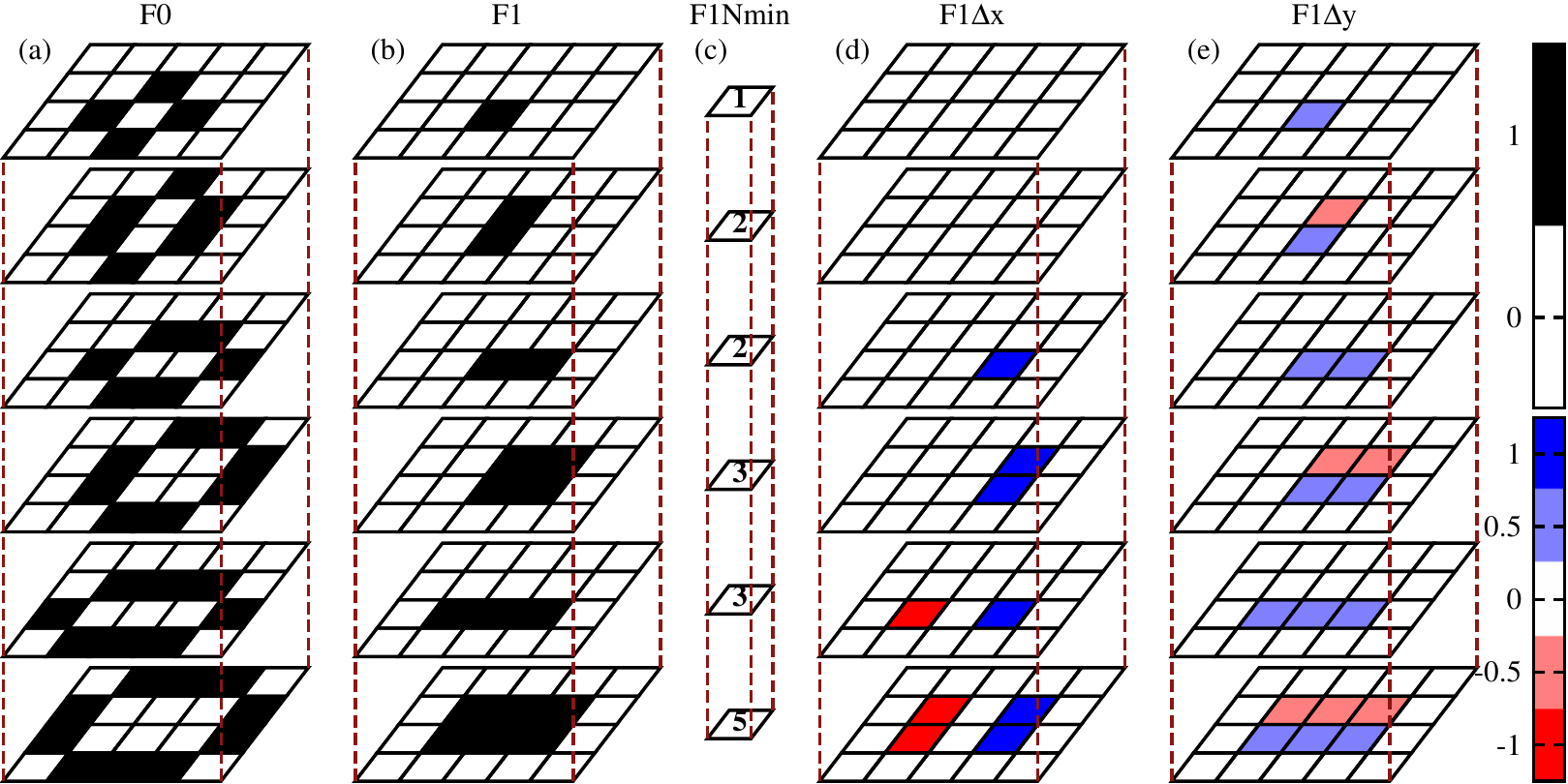}
  }
  \caption{Photon finding relies in matching the shape of the photon signal (distributed over one or more neighboring pixels) with one of the 6 features in filter F1 (b). The shape is identified by sufficient overlap with filter F1, as determined by the F1Nmin tensor (c), while not extended over the edge defined by filter F0 (a). The full convolutional model is shown in Fig.~\ref{fig:e}. Subpixel resolution can be obtained as shown in Fig.~\ref{fig:f}, using convolutional filters F1\textDelta x and F1\textDelta y depicted in (d) and (e), respectively. Typically only the first 4 features are used (we used 6 to match the charge sharing resulting from narrow \SI{25}{\micro\metre} pixels and underbiasing). F0, F1 and F1Nmin must be judiciously matched to detect all shapes of interest while preventing multiple feature matching.}
  \label{fig:d}
\end{figure}

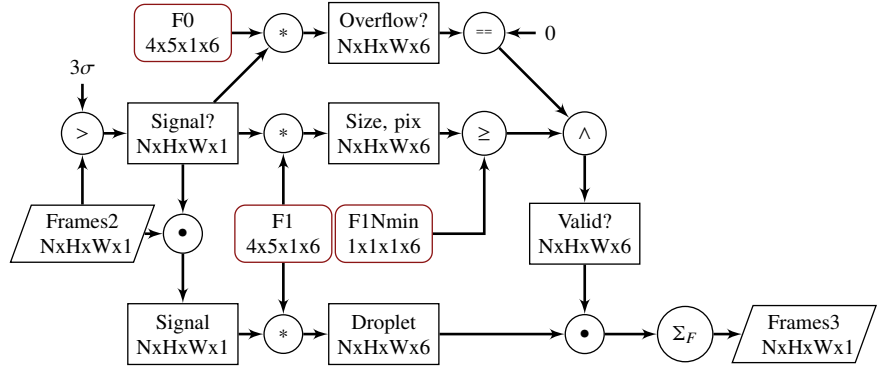
\begin{figure}[ht]
  %\centerline {%
    \begin{tikzpicture}[node distance = 5.2cm, auto]
            % Place nodes
            \node [io] (in)                         {Frames2\\NxHxWx1};
            \node [op, above of=raw] (op1)          {$>$};
            \node [param, above of=op1](noise)      {$3\sigma$};
            \node [block, right of=op1] (test)      {Signal?\\NxHxWx1};
            \node [op, right of=in] (op2)           {$\bullet$};
            \node [block, below of=op2] (insignal)  {Signal\\NxHxWx1};
            \node [weight, right of=op2](f1)        {F1\\4x5x1x6};
            \node [op, above of=f1] (op3)           {$\ast$};
            \node [op, below of=f1] (op4)           {$\ast$};
            \node [block, right of=op3](size)       {Size, pix\\NxHxWx6};
            \node [block, right of=op4](signal)     {Droplet\\NxHxWx6};
            \node [weight, above of=test](f0)       {F0\\4x5x1x6};
            \node [op, right of=f0] (op5)           {$\ast$};
            \node [block, right of=op5](overflow)   {Overflow?\\NxHxWx6};
            \node [op, right of=overflow] (op6)     {\tiny $==$};
            \node [param, right of=op6](zero)       {$0$};
            \node [weight, right of=f1](f1nmin)     {F1Nmin\\1x1x1x6};
            \node [op, right of=size] (op7)         {$\geq$};
            \node [op, right of=op7] (op8)          {$\land$};
            \node [block, below of=op8](valid)      {Valid?\\NxHxWx6};
            \node [op, below of=valid] (op9)        {$\bullet$};
            \node [op, right of=op9] (op10)         {\textSigma$_F$};
            \node [io, right of=op10](out)          {Frames3\\NxHxWx1};
            % Draw edges
            \path [line] (in) -- (op1);
            \path [line] (noise) -- (op1);
            \path [line] (op1) -- (test);
            \path [line] (in) -- (op2);
            \path [line] (test) -- (op2);
            \path [line] (op2) -- (insignal);
            \path [line] (test) -- (op3);
            \path [line] (f1) -- (op3);
            \path [line] (insignal) -- (op4);
            \path [line] (f1) -- (op4);
            \path [line] (op3) -- (size);
            \path [line] (op4) -- (signal);
            \path [line] (f0) -- (op5);
            \path [line] (test) -- (op5);
            \path [line] (op5) -- (overflow);
            \path [line] (overflow) -- (op6);
            \path [line] (zero) -- (op6);
            \path [line] (size) -- (op7);
            \path [line] (f1nmin) -| (op7);
            \path [line] (op6) -- (op8);
            \path [line] (op7) -- (op8);
            \path [line] (op8) -- (valid);
            \path [line] (signal) -- (op9);
            \path [line] (valid) -- (op9);
            \path [line] (op9) -- (op10);
            \path [line] (op10) -- (out);
    \end{tikzpicture} 
  %}
  \caption{Photon charge summing is based on several convolution operations ($\ast$) for validating photon hits and calculating the corresponding photon energy. The output is either the reconstructed photon energy placed in the appropriate pixel (for valid hits) or zero otherwise. Careful design of filters F0, F1 and F1Nmin allow detecting hits with arbitrary shapes in the different feature layers F while preventing multiple detection. This entire model is implemented in eight lines of Tensorflow code.}
  \label{fig:e}
\end{figure}

At low photon occupancy (sparse photons) and sufficient signal to noise ratio (typically the case for modern radiation imaging pixel detectors), individual photons can be measured. Their signal is often split over several pixels, leading to charge sharing and changes in the spectrum of deposited energy. To minimize the effect of charge sharing \cite{abboud2013sub}, first we match the different possible charge cloud shapes with the detected shape. Typical computer vision applications achieve this by using a convolution with matching filters represented by 4-rank tensors, as shown in Fig.~\ref{fig:d}: F0 and F1 have a shape of $[4,5,1,6]$, corresponding here to the \num{4x5} kernel that is used to convolve ($\ast$) at each pixel location; the dimension \num{1} must match the number of features of the input data set (in our case, \num{1}, see Fig.~\ref{fig:e}), and the last dimension is the number of features that we are extracting (i.e., the different images shown in each horizontal plane of F0 and F1).

Typical computer vision applications for feature segmentation (and object identification when using a deep learning model, i.e., a deep stack of convolutional layers) use a relatively fuzzy approach, with individual features obtained by training. Unlike in the deep learning approach, we designed two model branches with fixed weights (requiring no training and providing a clear physical meaning to each value) that strictly verify if the photon cloud shape conforms to one of the features in F1 while remaining within the boundaries defined by F0. Finally, the logical AND operation ($\land$) determines if there is a valid match. F0, F1 and F1Nmin must be chosen judiciously to ensure that any photon cloud shape will match only one feature, or none.

The top 4 layers of F0 and F1 unambiguously identify each of the 13 ``event types'' identified in \cite{abboud2013sub} (the key is the particular choice of values in test vector F1Nmin). With successful shape matching, the total charge of the photon cloud is assigned to the pixel with the highest signal (using the subpixel information extracted in next subsection to correctly assign the charge in the ambiguous case of even number of rows or columns for the charge cloud). The additional two layers elegantly extend the model to process charge clouds up to \num{2x3} pixels (and lead to a large increase in the number of event types), in order to process larger charge clouds. For typical pixel detectors, the first 4 features are sufficient.

This algorithm is mathematically equivalent with the one described in \cite{abboud2013sub}, which is also the state-of-the-art in charge summing with integrating detectors at LCLS. Our implementation is a generalization to charge clouds up to \num{2x3} pixel clusters and tests all possible charge cloud shapes at every pixel; the computation does not depend on the number of photons, and the maximum number of photons is returned. Typical droplet algorithms use an extra threshold parameter for detecting pixels where all the possible shapes are searched, thus missing some of the photons, and the computation time increases linearly with the number of photons; our model avoids both of these problems (while being 1 to 2 orders of magnitude faster).

Photon charge summing only works with sparse photons (typical images at LCLS-II); areas with high photon fluxes can be converted easily to photon counting images by division with the expected single photon gain followed by rounding to the nearest integer.

Tensorflow enables designing complex models with very compact and elegant code; we described the entire model in Fig.~\ref{fig:e} using only eight lines of code.

\subsection{Photon Subpixel Resolution}

\begin{figure}[ht]
  %\centerline {%
    \begin{tikzpicture}[node distance = 5.2cm, auto]
            \node [weight](f1dx)      {F1\textDelta x\\4x5x1x6};
            \node [block, above of=f1dx] (insignal) {Signal\\NxHxWx1};
            \node [op, right of=insignal] (op2)   {$\ast$};
            \node [block, right of=op2](mx)      {Momentum\\NxHxWx6};
            \node [op, right of=mx] (op3)   {$/$};
            \node [block, below of=mx](frames3)      {Droplet\\NxHxWx6};
            \node [block, right of=op3](xdev0)      {{ }\\NxHxWx6};
            \node [block, below of=xdev0](valid)      {Valid?\\NxHxWx6};
            \node [op, right of=xdev0] (op4)   {$\bullet$};
            \node [op, right of=op4] (op5)   {\textSigma$_F$};
            \node [io, right of=op5] (xdev) {\textDelta x\\NxHxWx1};
            % Draw Edges
            %\path [line] (f1) -- (op1);
            %\path [line] (dx) -- (op1);
            %\path [line] (op1) -- (f1dx);
            \path [line] (f1dx) -| (op2);
            \path [line] (insignal) -- (op2);
            \path [line] (op2) -- (mx);
            \path [line] (mx) -- (op3);
            \path [line] (frames3) -| (op3);
            \path [line] (op3) -- (xdev0);
            \path [line] (xdev0) -- (op4);
            \path [line] (valid) -| (op4);
            \path [line] (op4) -- (op5);
            \path [line] (op5) -- (xdev);
    \end{tikzpicture} 
  %}
  \caption{Extracting the \textDelta x subpixel photon centroids (expressed in pixel pitches) is straightforward and fast, reusing intermediary layers from Fig.~\ref{fig:e}. Similarly, \textDelta y can be obtained by using F1\textDelta y. The centroids require linearization to yield actual positions.}
  \label{fig:f}
\end{figure}
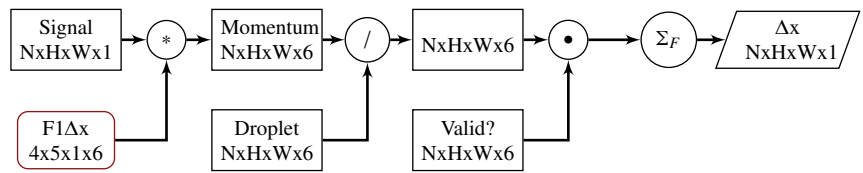

The subpixel centroid \textDelta x of photon charge clouds, measured in pixel pitches, can be easily calculated using the F1\textDelta x filters as shown in Fig.~\ref{fig:f}. The centroid position in the $y$ direction can be calculated similarly, just by using the F1\textDelta y filter. Note that the subpixel centroid has a nonlinear dependence on the actual position where the photon is detected.

This algorithm is mathematically equivalent to the ``classical'' algorithm described in \cite{cartier2014micron}, which is the state-of-the-art for subpixel resolution with integrating detectors at LCLS; our implementation includes a supplementary feature of zeroing the signal of pixels within noise,  resulting in increased signal-to-noise ratio for some photons.

\subsection{Summarization}
Tensorflow already has functions for histogramming, segmentation, sparsification, averaging and summing, etc. over arbitrary (combinations of) tensor dimensions; they are much faster than similar compiled functions, reducing massive data sets in real time and enabling online analysis with a minimum of overhead. Further functionality can involve, for example,  identifying diffraction spots in real time. Once the input data is reduced to either lists of sparse photons, or photon counting images, this data is much smaller, enabling faster computation and greatly increased compression rates for offline storage, without loss of photon information.

\section{RESULTS}
\subsection{Accuracy}
\begin{figure}[h]
  \centerline {%
    \includegraphics[width =6.5in]{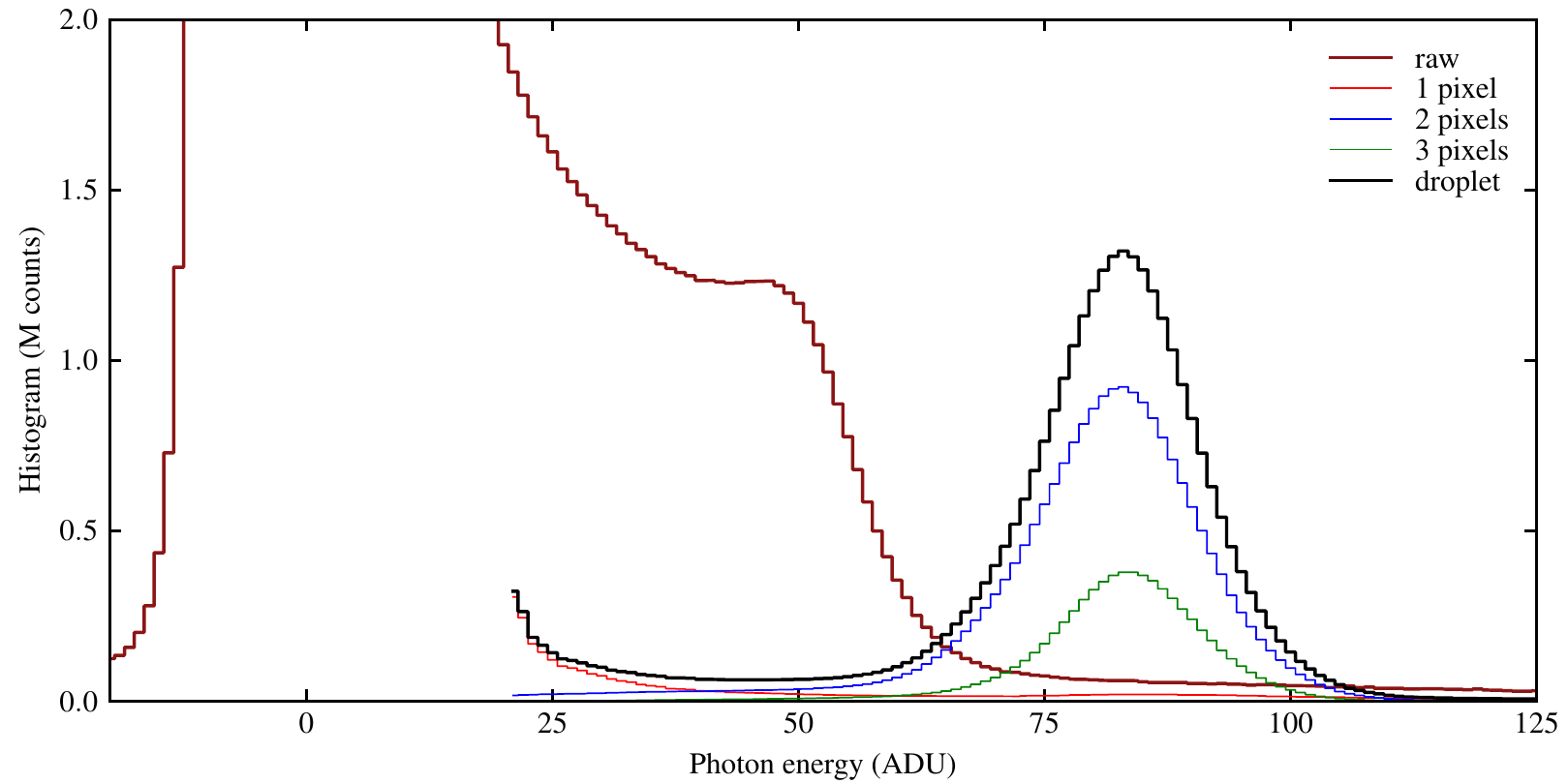}
  }
  \caption{Performance of photon charge summing model: the thick red line shows a histogram of Frames2 (before); the thick black line depicts a histogram of Frames3 (after charge summing). Note the radically improved spectrum (matching expected single photon gain of \num{83}~adu). The individual features obtained by summation over \num{1}, \num{2} or \num{3} pixels are indicated by thin lines. The system is dominated by \num{2} and \num{3} pixel events.}
  \label{fig:g}
\end{figure}

\begin{figure}[h]
  \centerline {%
    \includegraphics[width =6.5in]{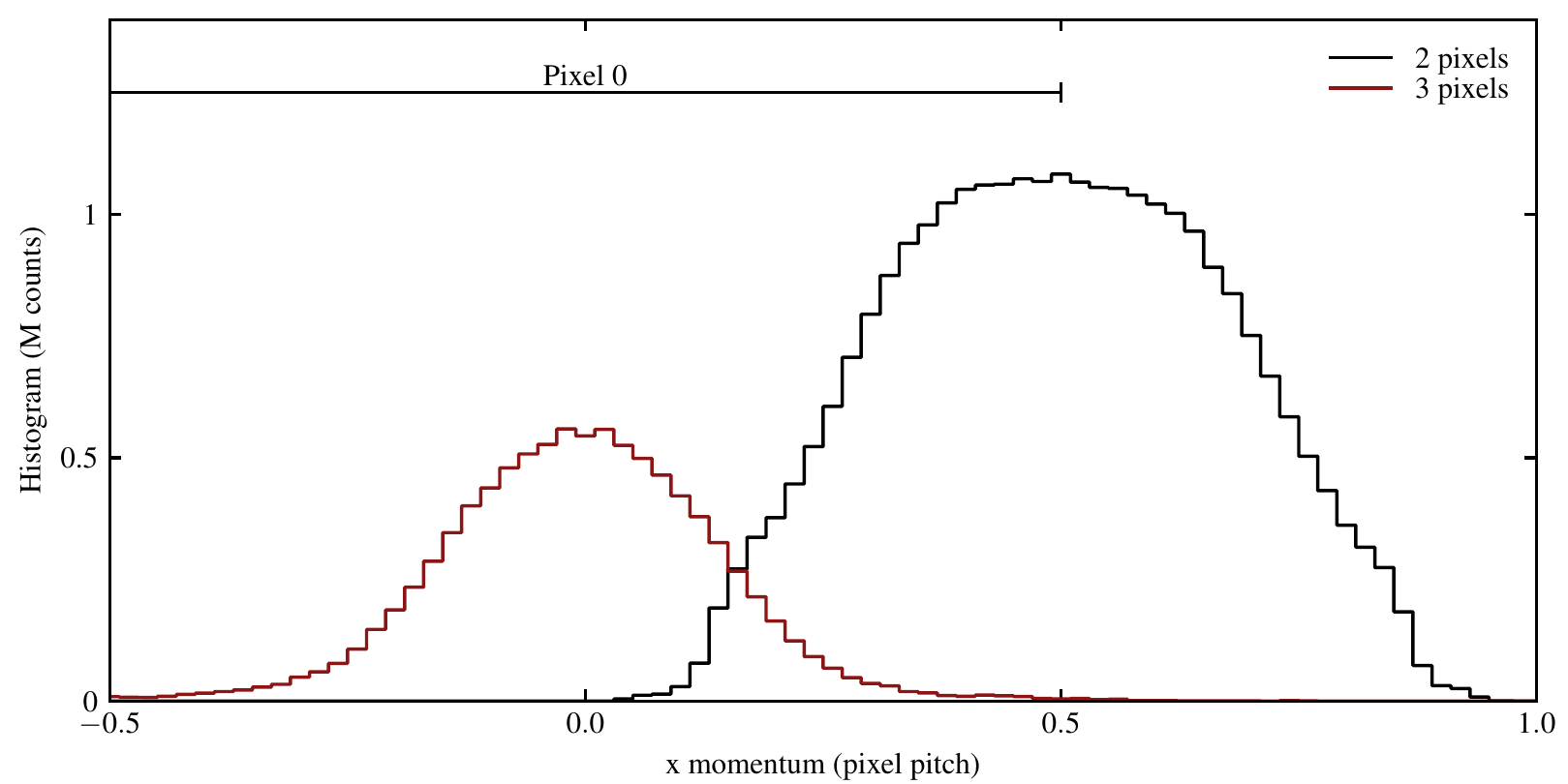}
  }
  \caption{Histogram of subpixel centroids in the $x$ direction; after linearizarion, photon position resolution in the order of \si{\micro\metre} is possible at low photon occupancy.}
  \label{fig:h}
\end{figure}

The ``classical'' CPU algorithms and novel Tensorflow algorithms yield exactly the same results, photon by photon. The only parameters that can be changed are the two noise thresholds shown in Fig.~\ref{fig:c} and Fig.~\ref{fig:e}, displayed as ``$3 \sigma$'' and typically set between $3 \sigma$ and $5 \sigma$ \cite{cartier2014micron}. Figure~\ref{fig:g} shows a spectrum obtained in (intentionally challenging) conditions: thick sensor (\SI{500}{\micro\metre}), underbiased at \SI{100}{\volt} (just over the minimum bias for full depletion, \SI{90}{\volt}), with a small pitch in one direction (\SI{25}{\micro\metre}) and relatively soft X-ray photons (\SI{5.9}{\kilo\electronvolt}). The thick red line shows the raw histogram, before charge summing; the black line shows the result of charge summing, with the peak at its expected location (\num{83}~adu for \SI{5.9}{\kilo\electronvolt} photons); the effect of charge sharing almost completely removed. Also the individual components with a charge cloud width of \num{1}, \num{2} and {3} pixels are shown with thinner lines. Note that most detected photons have charge cloud widths of 2 pixels, some have 3 pixels, and almost none have a cloud size of 1 pixel. In Fig.~\ref{fig:h} we show the subpixel centroids with the same detector settings (\SI{100}{\volt} bias). The majority of photon charge clouds are wider than \num{1} pixel, enabling recovery of subpixel position from subpixel centroids \cite{blaj2018hammerhead}.

\subsection{Performance}
\begin{figure}[h]
  \centerline {%
    \includegraphics[width =6.5in]{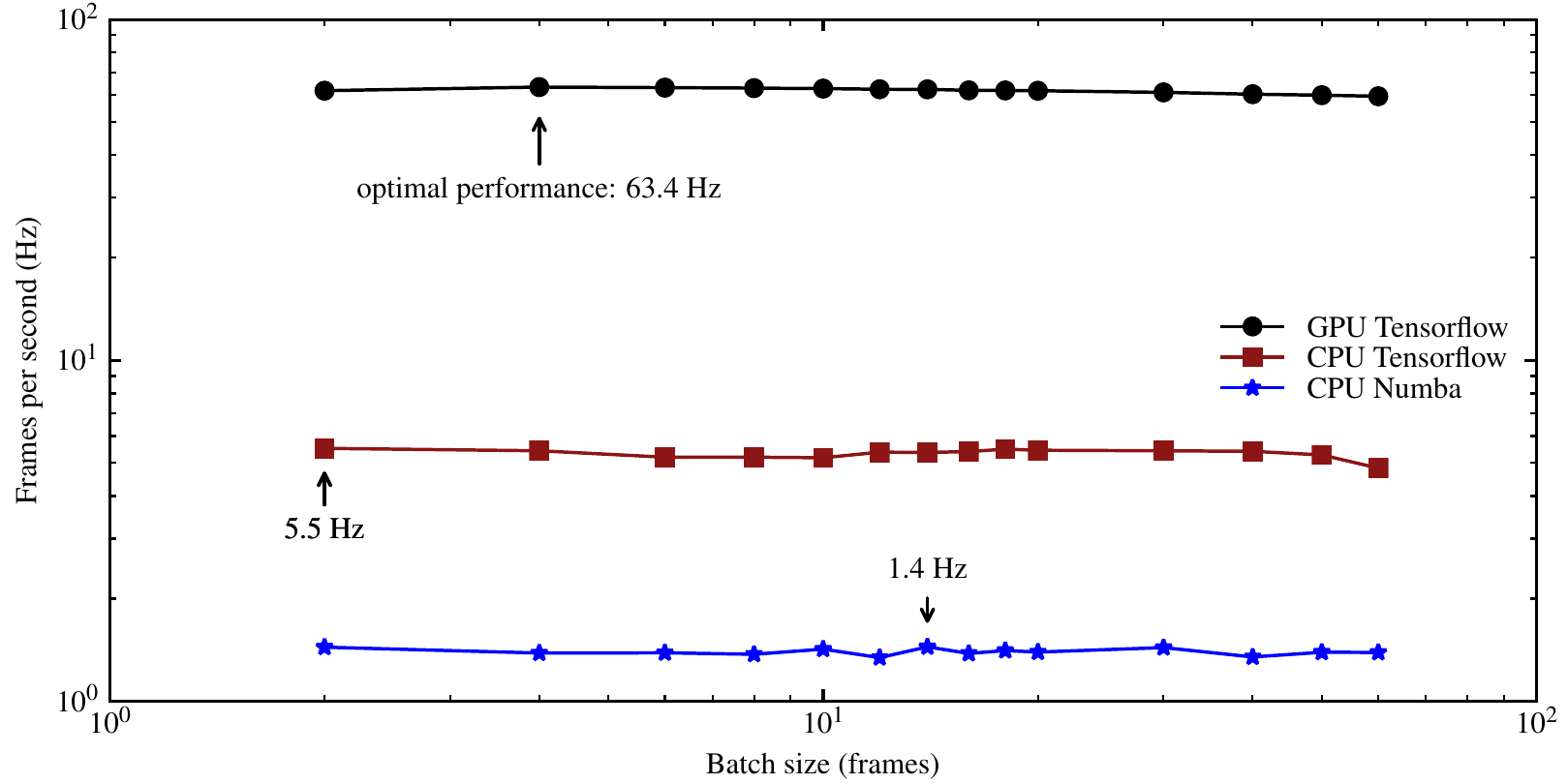}
  }
  \caption{Performance chart of the Tensorflow model, shown on a log-log plot. Blue stars indicate the performance of ``classical'' CPU compiled code (using Numba with careful performance optimizations) with a processing rate of \SI{1.4}{\hertz} (decreasing at higher occupancy, as each droplet is calculated separately). Red squares depict the same hardware running Tensorflow on the CPU at \SI{5.5}{\hertz}, or \num{4}x faster (note that higher occupancy does not increase computation time, as all features are calculated in parallel). Black dots show that the Tensorflow model is unleashed even by a modest consumer GPU, yielding \SI{63.4}{\hertz}, or a \num{46}x speed increase over the ``classical'' CPU code, while producing identical results.}
  \label{fig:i}
\end{figure}

To compare the performance of the novel Tensorflow model with previously developed Numba compiled code (complex and carefully optimized), we built a model consisting of: a dark and gain stage, 3 stacked common mode stages, a charge summing stage with 6 feature kernels, subpixel centroiding on both $x$ and $y$ direction and full summarization (photon sparsification, accumulating photon counting maps, histograms). The results are shown in Fig.~\ref{fig:i}.

Tensoflow running on a CPU was \num{4} times faster than the Numba code running on the same computer. However, running the Tensorflow model on a modest, inexpensive consumer GPU unleashed its performance, achieving a \num{46} times speed improvement over the ``classical'' CPU code.

The raw data of \num{50000} frames is about \SI{50}{GB} in size (about \SI{1}{MB} per frame); as it consists mostly of noise and relatively sparse photons, lossless compression yields about \SI{25}{GB}. However, extracting the full photon information (energy, coordinates with subpixel precision, cloud shape) and saving it into a file enables two orders of magnitude compression without loss of photon data. This compression ratio depends strongly on the statistics of individual FEL applications; most LCLS-II applications are expected to yield sparse data. Systematic tests with representative data sets are currently underway.

Writing the Tensorflow model required very little effort beyond the initial designing and testing work; currently it consists of a few tens of lines of code, due to the elegant, implicit handling of boundary conditions and power of calculating all possible features simultaneously.

\section{CONCLUSIONS}
Tensorflow code is compact and elegant, enabling extremely rapid development; it supports Python and Numpy natively; and it is simultaneously hardware agnostic and highly scalable. Last but not least, Tensorflow and other similar frameworks will continue to benefit from advances in machine learning hardware and software with minimal or no code changes. Faster GPUs, new tensor processing units, and other developments are likely to appear in the future, speeding up current models.

The performance of Tensorflow code running on a modest GPU now exceeds by far the execution speed of highly optimized compiled code, reaching \SI{63}{\hertz} now with ePix100 cameras, and over \SI{240}{\hertz} with ePix10K cameras (sufficient to process data from two ePix10K cameras in real time). While it might be possible to build even faster code using Nvidia CUDA, it would likely be difficult to develop and might depend strongly on the GPU architecture.% We are planning tests with more powerful GPUs and integration with data acquisition systems to enable real time processing of ePix100 data streams (with sustained speeds of \SI{120}{\hertz}).

Our novel Tensorflow model already reduces by 3 orders of magnitude the projected cost of online analysis and compression without photon loss at LCLS-II, and enables the development of a pipelined hardware system which is expected to yield an additional 3 to 4 orders of magnitude cost reduction for processing the LCLS-II pixel detector data at full speed (supporting online analysis, reduction to photon data and compression). This will turn a prohibitively expensive task into a merely challenging one.

Computer vision a decade ago was dominated by hand-crafted filters; their structure inspired the deep learning revolution resulting in modern deep convolutional networks. Similarly, while each layer of our novel hand-crafted Tensorflow filters has a clear meaning, the filters will provide a starting point for development of future deep learning models for ultrafast and efficient processing and classification of pixel detector images at FEL facilities. 

% Sections that will go in second font

% Acknowledgement
\section{ACKNOWLEDGMENTS}
Use of the Linac Coherent Light Source (LCLS), SLAC National Accelerator Laboratory, is supported by the U.S. Department of Energy, Office of Science, Office of Basic Energy Sciences under Contract No. DE-AC02-76SF00515. Publication number SLAC-PUB-17277.

% References

%\bibliographystyle{aipnum-cp}%
%\bibliography{main.bib}%

%Control: year  (1) truncated

\end{document}